\documentclass[conference]{IEEEtran}
\IEEEoverridecommandlockouts               % final (journal style)

\usepackage{macros}

\usepackage{mathptmx}                  % use matching math font
\usepackage{amsmath,amssymb,amsfonts}

\usepackage{multirow}
\usepackage{graphicx}
\usepackage{markdown}
\usepackage{todonotes}
\usepackage{subcaption}
\usepackage{textcomp}
\usepackage{xcolor}
\usepackage{cite}
\usepackage{hyperref}

\def\BibTeX{{\rm B\kern-.05em{\sc i\kern-.025em b}\kern-.08em
    T\kern-.1667em\lower.7ex\hbox{E}\kern-.125emX}}

\usepackage{fancyhdr}

\fancypagestyle{firstpage}{
  \fancyhf{}
  
  \fancyfoot[L]{\footnotesize \copyright\ 2026 IEEE. Personal use of this material is permitted. Permission from IEEE must be obtained for all other uses, in any current or future media, including reprinting/republishing this material for advertising or promotional purposes, creating new collective works, for resale or redistribution to servers or lists, or reuse of any copyrighted component of this work in other works.}
}

\graphicspath{{figs/}{figures/}{pictures/}{images/}{./}} % where to search for the images

\makeatletter
\newcommand{\linebreakand}{%
  \end{@IEEEauthorhalign}
  \hfill\mbox{}\par
  \mbox{}\hfill\begin{@IEEEauthorhalign}
}
\makeatother

\begin{document}

%% Paper title.
\title{From Docket Minutes to Case Trajectories: Building Analysis-Ready Criminal Court Data}

%% Author ORCID IDs should be specified using \authororcid like below inside
%% of the \author command. ORCID IDs can be registered at https://orcid.org/.
%% Include only the 16-digit dashed ID.
% \author{%
%   \authororcid{Erfan {Hosseini Sereshgi}}{0000-0003-2548-7428},
%   \authororcid{Kory Rosen}{0009-0008-7763-9765}
%   Jaya Dhariwal,
%   \authororcid{Nicholas Mattei}{0000-0002-3569-4335}, and 
%   \authororcid{Aron Culotta}{0000-0003-2660-7575}
% }
\author{\IEEEauthorblockN{Erfan {Hosseini Sereshgi}}
\IEEEauthorblockA{\textit{The Murphy Institute} \\
\textit{Tulane University}\\
New Orleans, LA, USA \\
shosseinisereshgi@tulane.edu}
\and
\IEEEauthorblockN{Kory Rosen}
\IEEEauthorblockA{\textit{Tulane University}\\
New Orleans, LA, USA \\
krosen5@tulane.edu}
\and
\IEEEauthorblockN{Jaya Dhariwal}
\IEEEauthorblockA{\textit{Tulane University}\\
New Orleans, LA, USA \\
jdhariwal@tulane.edu}
\linebreakand
\IEEEauthorblockN{Nicholas Mattei}
\IEEEauthorblockA{\textit{Tulane University}\\
New Orleans, LA, USA \\
nsmattei@tulane.edu}
\and
\IEEEauthorblockN{Aron Culotta}
\IEEEauthorblockA{\textit{Tulane University}\\
New Orleans, LA, USA \\
aculotta@tulane.edu}
}

%%%%%%%%%%%%%%%%%%%%%%%%%%%%%%%%%%%%%%%%%%%%%%%%%%%%%%%%%%%%%%%%
%%%%%%%%%%%%%%%%%%%%%% START OF THE PAPER %%%%%%%%%%%%%%%%%%%%%%
%%%%%%%%%%%%%%%%%%%%%%%%%%%%%%%%%%%%%%%%%%%%%%%%%%%%%%%%%%%%%%%%
\maketitle

\thispagestyle{firstpage}
% Abstract section.
\begin{abstract}%
 % filler text. Replace with your abstract.
  %
  %% We recommend that you link to your supplemental material here in the abstract, as well
  %% as in the Supplemental Materials section at the end.
  %The decentralization and structural complexity of the United States judicial system make criminal court data notoriously difficult to access, interpret, and visualize. Judicial records are predominantly unstructured, multidimensional, nonlinear and distributed among different organizations, creating a barrier to transparency and public oversight. Overcoming this data opacity requires tools accessible to a diverse audience, ranging from legal professionals to the general public. In this paper, we introduce an interactive, web-based visual analysis dashboard designed for New Orleans criminal court data. Our system provides intuitive visualization techniques, based on experts' and community feedback, to analyze demographics, temporal trends, and specific case events, enabling users to identify anomalies, evaluate the effects of policy changes, and detect procedural irregularities. We demonstrate the system's design and capabilities through an exploratory analysis of recent short-term trends and policy impacts within New Orleans. Furthermore, we examine the effectiveness and usefulness of the proposed system through a small-scale formative study with community stakeholders. 

    Criminal court records are often fragmented, noisy, and siloed, making it challenging for both legal experts and community stakeholders to access and analyze court data to identify trends and provide public oversight. In this paper, we introduce a system for data ingestion, harmonization, and dashboard visualization designed for New Orleans criminal court proceedings data, aimed at both expert and lay audiences. Our system collects and harmonizes disparate data, generates case trajectories, and provides intuitive visualization techniques based on community feedback to analyze demographics, temporal trends, and specific case events. We demonstrate the system through an exploratory analysis of recent short-term trends and policy impacts within New Orleans, illustrating how it can support the investigation of anomalies, policy changes, and procedural irregularities. We also present feedback from a small formative study of community stakeholders who have helped co-design and use the system, suggesting that it may improve access to and interpretation of local court data.
\end{abstract}


%% \section{Introduction} %for journal use above \firstsection{..} instead
\section{Introduction}
\label{sec:intro}
The criminal legal system operates through a sequence of high-stakes decisions, including charging, bail setting, plea offers, and sentencing, made under heavy caseloads and variable local rules. Despite the obvious importance of this system, very few jurisdictions have access to real-time, high quality data to inform policy making and day-to-day decision-making.  For example, fewer than half of U.S. prosecutors' offices collect sufficient data to report even basic statistics on bail, charging, and disposition~\cite{olsen2018prosecutors}.

This need is particularly urgent in New Orleans, which exhibits among the highest incarceration and exoneration rates in the United States~\cite{nre2019exonerations}. Court personnel and reform advocates wrestle with ongoing debates around cash bail inequities, charge refusal rates, pretrial detention, and case delays~\cite{digard2019justice,aclula2020pretrial,armstrong2022louisiana}, yet lack reliable data to inform debate.
% Answering these questions requires a unified, case-centric data system rather than aggregated summaries that mask underlying procedural trajectories.

% These decisions profoundly affect housing, employment, and public budgets long before a verdict, driving demand for data-driven court oversight~\cite{armstrong2022louisiana}. This need is particularly urgent in New Orleans, which exhibits among the highest incarceration and exoneration rates in the United States~\cite{nre2019exonerations}. 

% The consequences of this data gap are immediate. Pretrial detention decisions occur shortly after arrest, with median bail amounts approaching household median income in Louisiana, where many detainees face non-violent charges~\cite{aclula2020pretrial}. Long statutory charging windows allow prolonged detention that disrupts lives even when cases are ultimately declined~\cite{digard2019justice}. Organizations like Court Watch NOLA require case-level data to evaluate disparities in bail amounts, charging delays, case durations, and charge stacking across judicial sections, offense types, demographics, and temporal periods. Answering these questions requires a unified, case-centric data table rather than aggregated summaries that mask underlying procedural trajectories.

% Furthermore, institutional tracking remains limited; fewer than half of U.S. prosecutors' offices collect sufficient data to report basic statistics on bail, charging, and disposition~\cite{olsen2018prosecutors}.

For community organizations in Orleans Parish, public court records are fragmented, unstructured, and difficult to consolidate. Defendants typically enter through Magistrate Court for bond and probable cause hearings before transferring to Criminal District Court upon formal charging, complicating data analysis. 
% removed bit about non-linking case IDs; those do exist.
While a simple public access portal does exist, it provides only individual case queries without structured exports or APIs, making large scale analysis cumbersome. Furthermore, important case events are denoted mostly through clerk notes, requiring information extraction from noisy and jargon filled free text data.

% returning heterogeneous data that mixes demographic fields, charge rows, and abbreviated clerk minutes. Varied shorthand, parallel entries, and partial charge acceptances create significant technical hurdles for real-time reform advocacy.

% The consequences of this data gap are immediate. Pretrial detention decisions occur shortly after arrest, with median bail amounts approaching household median income in Louisiana, where many detainees face non-violent charges~\cite{aclula2020pretrial}. Long statutory charging windows allow prolonged detention that disrupts lives even when cases are ultimately declined~\cite{digard2019justice}. Organizations like Court Watch NOLA require case-level data to evaluate disparities in bail amounts, charging delays, case durations, and charge stacking across judicial sections, offense types, demographics, and temporal periods. Answering these questions requires a unified, case-centric data table rather than aggregated summaries that mask underlying procedural trajectories.

\paragraph*{\textbf{Contributions}}
To address this gap, we present a data ingestion, record linkage, and classification pipeline paired with an interactive public dashboard for Orleans Parish in collaboration with legal advocates, researchers, and journalists. Our system ingests public docket records, joins siloed magistrate and district court identifiers, extracts procedural events from clerk text, and reconstructs analysis-ready case trajectories. Specifically, we (i) collect records and link magistrate dockets to criminal court cases using arrest cross-references and minute text citations; (ii) apply quality filters to isolate completed single-defendant cases; (iii) map clerk notes to twenty-one procedural labels using jurisdiction-tuned regular expressions; (iv) model case progression on an explicit state graph; and (v) identify similar cases for a given case. We deploy these data within an interactive dashboard,
%featuring coordinated filters
enabling community stakeholders to analyze systemic patterns in detention, delay, and disposition.

Using this dashboard, we investigate two years of criminal court data, demonstrating how the system can be used to investigate how case length, charge refusal rates, and bond amounts vary by defendant, charge, and time. Furthermore, we show how case similarity measures can be used to identify clusters of similar trajectories. While our case study demonstrates the system for one jurisdiction, the general methodology can be adapted to other jurisdictions with fragmented court data to expand public oversight of criminal court.
%The pursuit of transparency remains a critical imperative within the criminal justice system, particularly in jurisdictions like New Orleans \cite{armstrong2022louisiana} (though New Orleans is not uniquely lacking or an outlier for mid-sized cities in the USA). On the national scale, recent initiatives have sought to mitigate systemic opacity and enhance institutional accountability through targeted reforms and data-driven oversight mechanisms. The application of visualization techniques to criminal case data spans multiple levels of abstraction, ranging from micro-level analysis of digital forensic evidence to macro-level management of judicial backlogs and public transparency. 
%Throughout this paper we may refer to the criminal district court as \emph{CDC} and the magistrate court as \emph{MAG}. 

\section{Related Work}
\label{sec:relatedwork}
Transforming fragmented criminal court records into analysis-ready tables is fundamentally a data engineering challenge, defined by disparate sources, missing relational joins, unstructured text, and systematic quality defects. Downstream tools, including visualizations and dashboards, rely on this structured foundation to make legal records transparent and accessible for public oversight. This section synthesizes prior research across legal information extraction, timeline modeling, exploratory quality checking, event-sequence representations, and public court-data platforms.%, thereby positioning our contributions within the broader literature.

\subsection{Visualizing Event Sequences and Timelines}
\label{sec:rw-sequences}

Criminal cases are inherently temporal, consisting of discrete proceedings that unfold over time. Visualizing these sequences presents challenges due to high dimensionality and temporal variance. Zinat et al.\ propose a task framework categorizing sequence analysis into distinct intents and strategies~\cite{zinat2024multilevel}, while Ruddle et al.\ aggregate event types into ``Event Maps'' to summarize procedural frequencies~\cite{ruddle2016methods}. We combine aggregation (Sankey case-flow diagrams) and segmentation (case state timelines), applying them to a jurisdiction-specific procedural graph so stakeholders can examine how dockets actively progress through magistrate and district courts.

Visualizing temporal and multivariate attributes simultaneously remains difficult. Montana et al.\ introduced ``EventBox'' to display duration, start times, and categorical variables concurrently~\cite{montana2025eventbox}, and Ortigossa et al.\ show that multidimensional time series require dimensionality reduction with coordinated views~\cite{ortigossa2025timeseries}. However, EventBox targets non-judicial encodings, while standard coordinated views do not address cross-court record linkage. Our dashboard maintains attributes like race, charges, and bail amounts as global filters across linked views rather than compacting them into single glyphs, preserving delay and outcome distributions for all users.

For large-scale event sequences, IVESA scales via sequence clustering and motif substitution~\cite{bernard2024ivesa}, while LIVE-ITS retrieves representative time series using LSH indexing~\cite{chen2024liveits}. Rather than optimizing over generic numeric streams, we address a smaller, noisier court corpus by combining discrete procedural labels, an explicit case-state graph robust to skipped hearings, and nearest-neighbor retrieval across harmonized identifiers.

\subsection{Exploratory Data Analysis and Quality Assurance}
\label{sec:rw-eda}

Visual inspection is an essential preliminary step for data quality verification, surfacing coding anomalies and other issues~\cite{maltz2010look}. Because Orleans Parish minute entries are abbreviated, inconsistent, and split across two numbering systems, our ingestion filters, rule-based labelers, and date-versus-elapsed-time scatterplots are explicitly structured to expose missing events, stalled dockets, and clerical artifacts \emph{before} stakeholders evaluate demographic or policy trends.

\subsection{Legal Information Extraction and Timeline Intelligence}
\label{sec:rw-legalai}

Before case sequences can be rendered visually, procedural facts must be extracted from unstructured judicial text. Datasets like LEEC benchmark element extraction for sentencing factors~\cite{zongyue2023leec}, and LLM evaluations demonstrate strong entity recognition across corpora~\cite{deusser2024comparative}. However, LEEC targets Chinese judgments, and LLM entity recognition remains vulnerable to prompt sensitivity and boundary errors. We instead employ twenty-one localized procedural labels extracted via deterministic regular expressions, prioritizing inspectable classification logic over open-ended coverage.

Beyond entity extraction, legal analysis requires structured event chronologies and outcome estimations. LLMs can summarize timelines but frequently misorder distant temporal events~\cite{mao2024comparative}, while judgment prediction models estimate penalty severity have been explored using hierarchical attention~\cite{liu2022penalty} and transformer models have predicted next proceeding outcomes~\cite{rosen2026arraignment}. We reconstruct chronologies directly from dated docket entries to eliminate date hallucinations, focusing on empirical state durations, detention statuses, and demographic variations rather than forecasting outcomes.

\subsection{Dashboards for Transparency and Decision Making}
\label{sec:rw-dashboards}

Visual dashboards are established tools for Evidence-Based Decision Making~\cite{Sarikaya_dashboard2019}. Prior systems track jail populations~\cite{riley2015criminal}, prioritize court backlogs via complexity scores~\cite{bhoir2025court}, or publish open prosecutorial data~\cite{elbashir2022improving}. However, municipal portals rely on pre-structured agency exports lacking cross-court linkage, while operational systems optimize internal backlog rather than public accountability. We scrape and join disparate magistrate and district court records to expose charge stacking, bail distributions, and section delays and other issues identified as critical by our community partners in an accessible interface.

Addressing legal information access, Amato et al.\ advocate for federated search with RAG for distributed sources~\cite{amato2024optimizing}, and Bakir et al.\ propose model-based metrics aligned with human judgment~\cite{bakir2023developing}. Because our deployment uses public Louisiana dockets, federated retrieval is unnecessary; we make the linked public record directly explorable while ensuring all labels, filters, and state transitions remain transparent.

In domain-specific visual analytics, systems like Visilant~\cite{zakopcanova_visilant} and CrimAnalyzer~\cite{garciazanabria_crimanalyzer} serve law enforcement, whereas Artemus~\cite{roberts_artemus} guides citizens through administrative navigation. Our dashboard bridges these roles, supporting trend and anomaly investigation for experts and community advocates around the harmonized court case.

\begin{figure*}[ht]
    \centering
    \includegraphics[width=\linewidth]{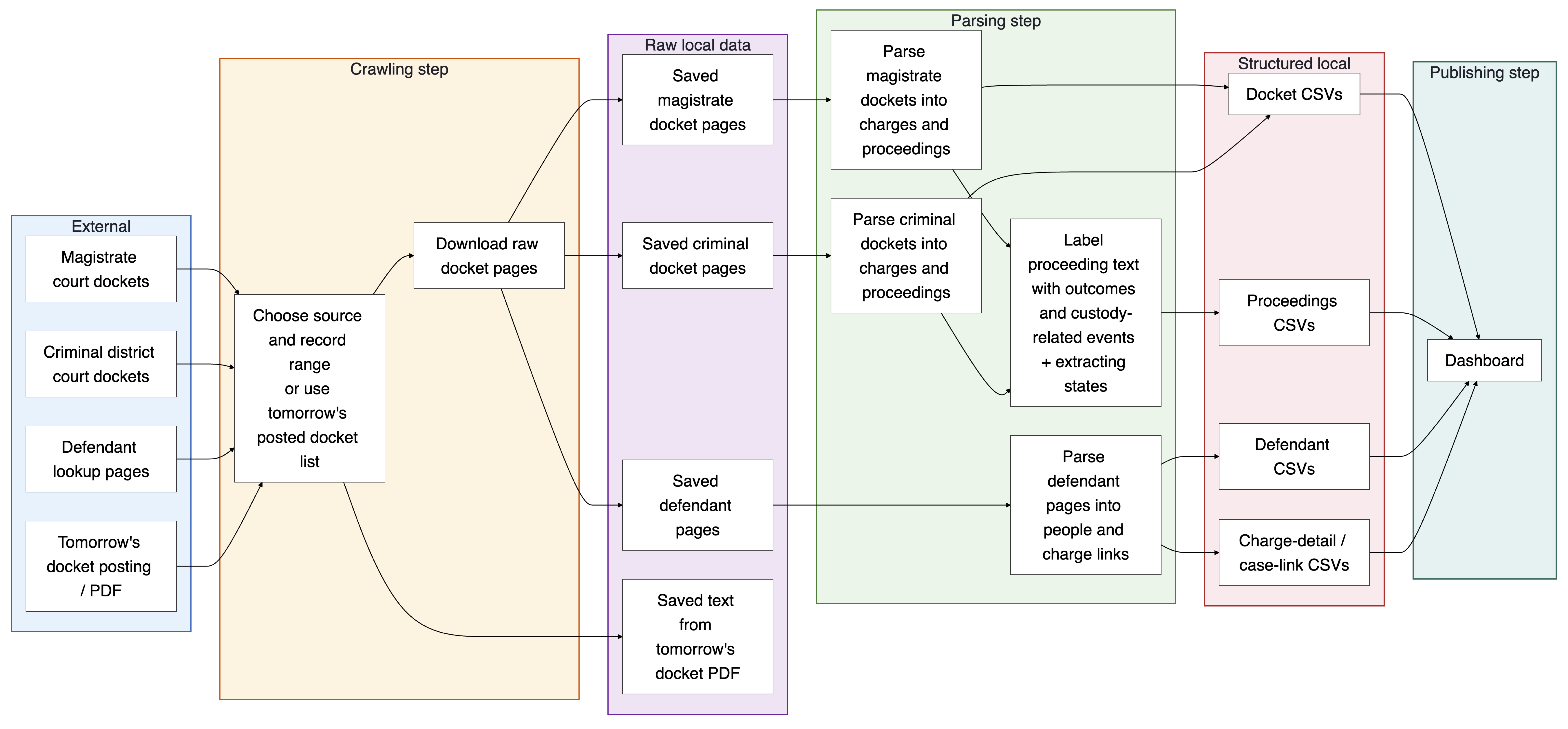}
    \caption{Data flow diagram for data acquisition, extraction, harmonization, and visualization.}
    \label{fig:data-flow}
\end{figure*}

\section{Methodology}
\label{sec:method}
Our system comprises seven main steps: \secref{data-acquisition} and \secref{data} detail the origin and format of the data, including the extraction and pre-processing protocols. \secref{label} describes the case event labeling system, which applies regular expressions to identify significant case events from the clerk notes for each proceedings (event). \secref{states} outlines the logic used to determine the state of a case based on the generated labels and other attributes. \secref{pretrial} demonstrates the process of identifying cases with pre-trial detainment and its duration. \secref{similar} presents our approach for finding similar cases.
Finally, \secref{vis} presents the visualization framework, highlighting the specific events and variables it surfaces alongside comparative examples from the criminal justice literature. \figref{data-flow} shows an overview of our system.

\subsection{Data Acquisition}
\label{sec:data-acquisition}
%\aron{talk about courts-etl process and why it's hard; e.g., crawling dockets}

%Criminal cases in New Orleans may pass through one or both of two court divisions that use entirely separate case identifiers, with no built-in way to link a magistrate record to its possible corresponding Criminal District Court case. The Orleans Parish Magistrate Court handles first appearances, bond settings, and probable cause determinations, while the District Attorney decides whether to accept or refuse charges. If charges are accepted and a Bill of Information is filed, the case is transferred to the Orleans Parish Criminal District Court, which presides over pre-trial proceedings, plea negotiations, and trial. Both court divisions were served by the Docket Master portal from the Orleans Parish Sheriff's Office, which is public under the Louisiana Public Records Act (R.S. 44:1 et seq.). \kory{However, the Docket Master did not offer any structured export, API system, or structured list indices of valid record numbers.} As a result, our dataset reflects only a limited subset of the records available through Docket Master, specifically those captured through our scraping process.
Criminal proceedings in New Orleans span two distinct court divisions with disconnected case-numbering systems: Magistrate Court (initial appearance, bail, and probable cause) and Criminal District Court (formal prosecution, plea negotiations, and trial). Connecting a magistrate record to its district court continuation requires custom record linkage, no cross-system identifier is publicly accessible. Public records from both divisions were hosted on the Sheriff's Office Docket Master portal (governed by Louisiana R.S.~44:1 et seq.).

\begin{figure}[t]
    \centering
    \includegraphics[width=\linewidth]{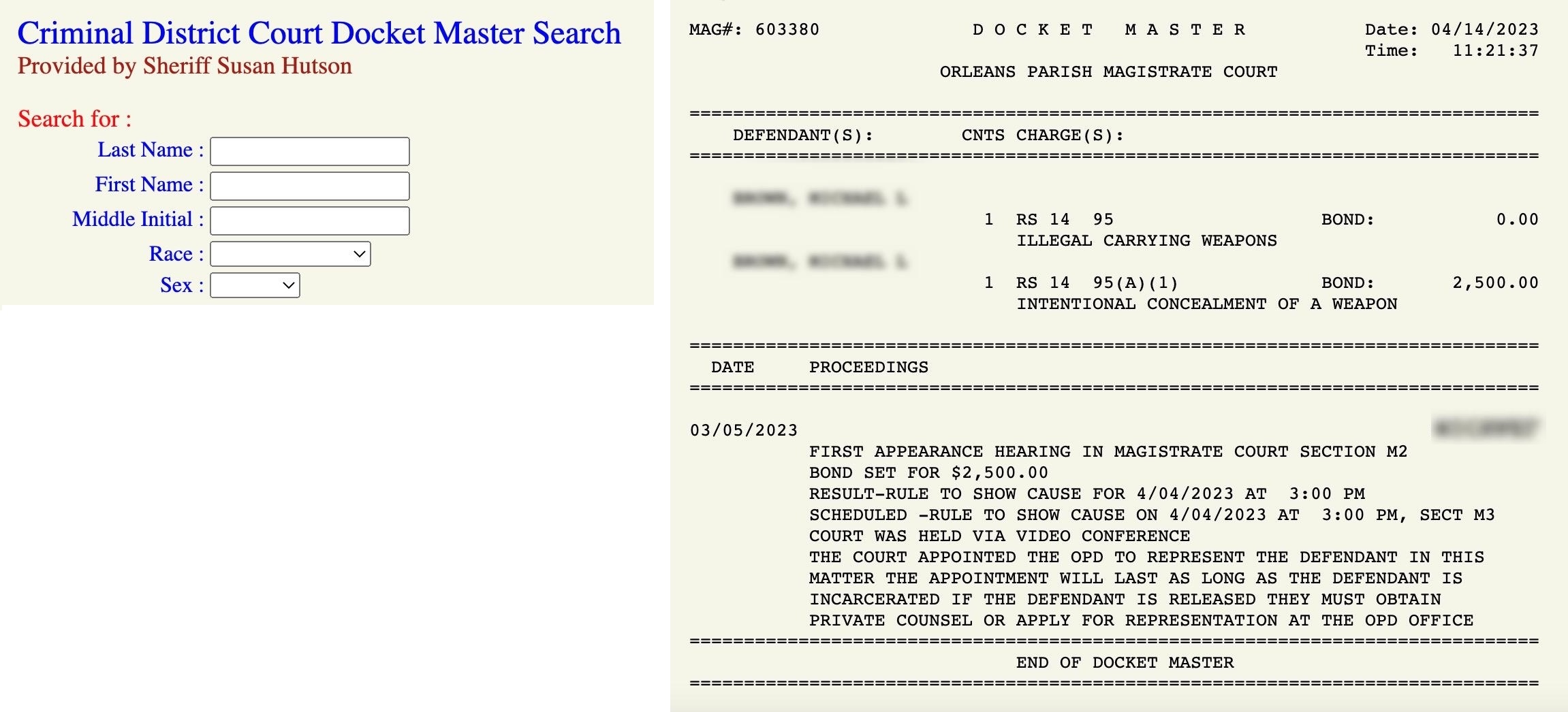}
    \caption{Docket Master landing page (left) and a sample docket page (right).}
    \label{fig:docketsample}
\end{figure}

Acquiring these records at scale presented a data engineering challenge rather than a simple database transfer. Docket Master lacked APIs, structured exports, or a searchable case index; interactive queries required exact identifiers and returned disconnected HTML pages for the same arrest. Each record yielded an unformatted web report containing defendant demographics, charge tables, and chronological clerk entries designed solely for human viewing (See \figref{docketsample}).

We developed an automated daily scraper that discovers candidate identifiers, revisits active dockets, and writes raw HTML reports to a staging repository. Consequently, our dataset represents a systematically captured subset of public dockets rather than an official exhaustive archive.

%\kory{write part about }

For each case, our custom web scraper gathers three specific record types from the Docket Master:

\begin{itemize}
    \item \textbf{Defendant profiles}, storing name, race, sex, and birth year
    \item \textbf{Charge entries}, capturing the statutory code, charge text, and bond amount
    \item \textbf{Chronological proceedings}, recording the date, presiding section, and the raw text of each clerk entry
\end{itemize}

% see https://sse.tulane.edu/sites/default/files/2023-10/cwnola_capstone_2022.pdf

\subsection{Data Ingestion and Harmonization}
\label{sec:data}

Since the two court divisions use unique case identifiers, cases are linked across them using two features: the arrest charge records of the defendant, which reference a magistrate number and a CDC case identifier; and the text of a CDC proceeding, where magistrate numbers could be embedded. Then, these are joined into a unified case-level table.

We then apply numerous sequential filters to yield a clean dataset. First, we retain only cases with complete defendant information. Second, we retain only single-defendant cases, as multi-defendant cases combine labels across defendants for a given proceeding and cannot be disambiguated from each other. Third, we retain only magistrate-started cases. Fourth, we exclude extradition cases, as defendants held for or transferred to out-of-state jurisdictions follow a trajectory that is not representative of most cases. Lastly, we remove proceedings whose texts contain no detectable legal event, as these are for scheduling or administrative purposes. %Lastly, we keep only completed cases, which are cases that have reached a terminal outcome (i.e., charge dismissal, guilty plea, verdict reached, or sentencing) that contain at least two labeled proceedings. 
%\erfan{Was this the case with the previous version of the data? Because I actually have a case closure detection method}
%\erfan{Kory+Aron?}

%\nick{Erfan to add here a thing like 125,000 magistrate cases covering XXX years with YYY entries, ZZZ charges, WWW unique defendnets, ETC...}
This resulted in the extraction of $130,640$ cases including $125,588$ MAG cases and $27,521$ CDC cases from 1992-10-26 to 2026-08-20 consisting of $820,564$ proceedings and $2,137$ unique charge names. %$47.8\%$ of these cases are able to be verified as closed from the clerk notes. %\aron{right? otherwise more than half should be closed...} \erfan{I got these numbers from Kory, I am not sure how he identifies a case is closed since he does not have a dedicated label for that. asking him now}

\subsection{Label Generation}
\label{sec:label}

Each proceeding entry in the Docket Master is a court minute entry, the official record of what occurred at a given hearing, prepared by a clerk. These entries follow no standardized format or controlled vocabulary, with abbreviations and shorthand notation varying from clerk to clerk. An example can be seen in \figref{docketsample}. As a result, the same legal event may appear in dozens of different phrasings depending on which clerk wrote it and when. For example, a probable cause finding can appear as \texttt{THE COURT HAS FOUND PROBABLE CAUSE}, \texttt{CFPRC}, \texttt{PROB CAUSE FOR}, and \texttt{P.C}. %\erfan{might be good to also mention typos and unfinished thoughts}

Let $\mathcal{L}$ be the set of twenty one labels representing key procedural events and outcomes across the full case lifecycle:

\[
\mathcal{L} = \left\{
\parbox{0.85\columnwidth}{\raggedright \small
  First Appearance Hearing, Bond Set, Bond Not Set,
  Released on Own Recognizance, Public Defender Appointed,
  Ankle Monitoring Imposed, Domestic Stay-Away Order Issued,
  Charges Accepted, Charges Refused,
  Probable Cause Found, Probable Cause Not Found,
  Defendant in Custody and Not Present in Court, State Not Present,
  Pre-Trial Conference Held, Continuance Occurred,
  Guilty Plea Entered, Not Guilty Plea Entered, Found Not Guilty,
  Verdict Returned at Trial, Sentence with Incarceration Imposed,
  Sentence Without Incarceration Imposed
}
\right\}
\]

For a proceeding $p$, let $L(p) \subseteq \mathcal{L}$ be its label set. A label $\ell \in \mathcal{L}$ is included in $L(p)$ if and only if the text of $p$ matches its associated regular expression. Since a single clerk entry may record multiple events, $L(p)$ may contain multiple labels. 

We implement this labeling process as a rule-based extraction library comprising $15$ extractor functions that, together, generate all $21$ labels in $\mathcal{L}$. Rather than mapping one function to one label, several extractors evaluate multiple outcomes. For example, the bond extractor evaluates whether bond was set or not, and the plea extractor determines whether the plea was guilty or not guilty. Each extractor matches the observed phrasings for its targeted event. More complex labels also incorporate exclusion rules to filter out false positives arising from scheduling-related language and spurious matches. Some extractors directly embed lists of Louisiana correctional facilities and parishes into their compiled patterns, since clerks typically refer to specific institutions or jurisdictions rather than using generic custody language. 

Each proceeding is processed by all extractors, and the resulting labels are deduplicated to form $L(p)$.  We chose this approach because the abbreviated, inconsistent format and jurisdiction-specific language of these entries make the application of off-the-shelf pretrained models unreliable without major task-specific tuning. 

We conducted an iterative process to refine extraction, inspecting the output to identify errors as well as reviewing unlabeled entries to improve recall. In the end, we sampled 100 entries for manual validation, finding no extraction errors in the sample.

\subsection{Identifying Case States}
\label{sec:states}
\begin{figure}[t]
    \centering
    \includegraphics[width=0.9\linewidth]{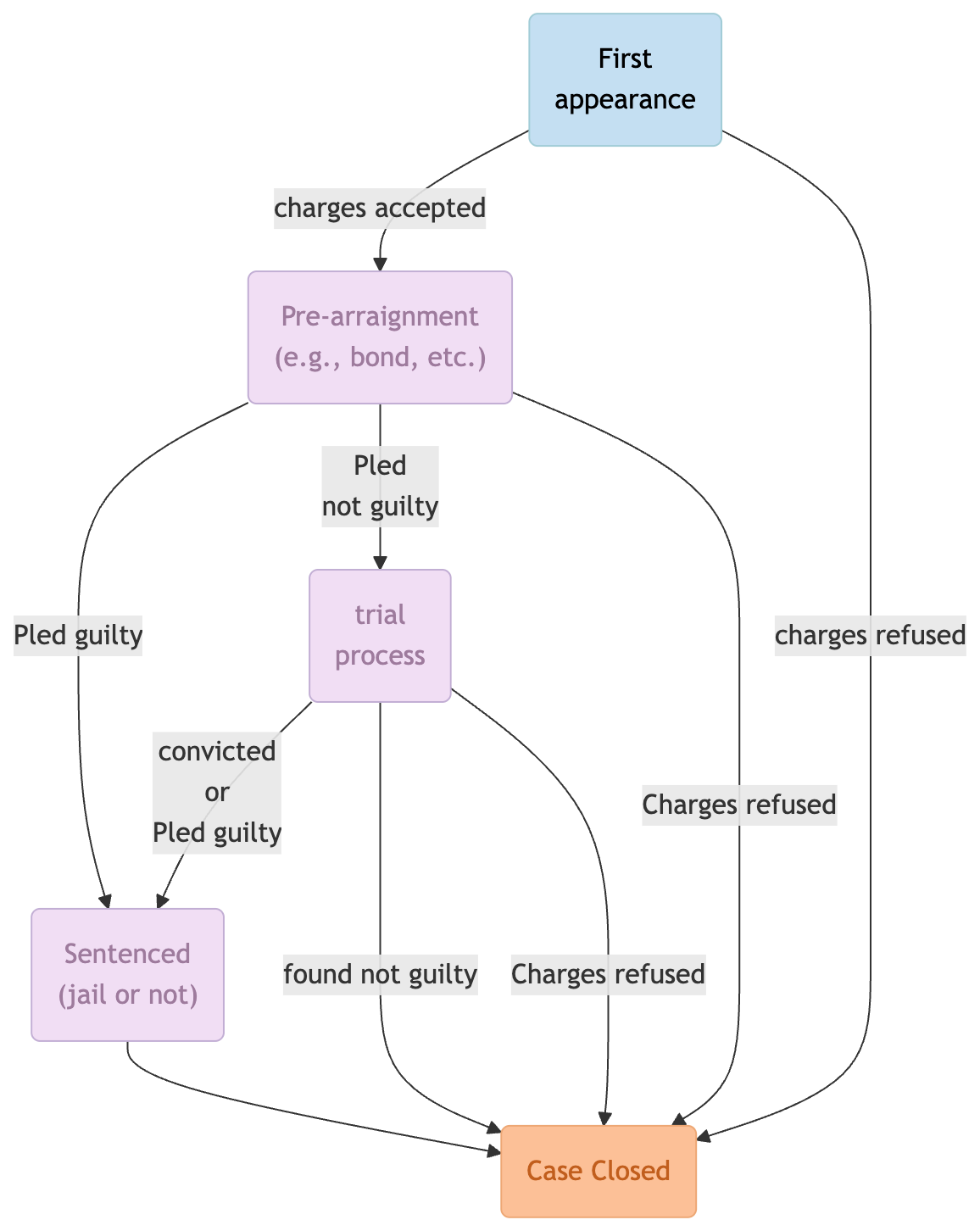}
    \caption{Flow chart of criminal cases. }
    \label{fig:flowchart}
\end{figure}

Initially, we establish a flow system representing the judicial process, which resembles the structural framework outlined in the comprehensive flowchart provided by the U.S. Department of Justice~\cite{bjs_justice_system_2021}. \figref{flowchart} demonstrates the high-level states of a criminal case, spanning from First appearance to Case closure. While certain labels are used strictly for filtering purposes, others serve as critical indicators for pinpointing a case's exact trajectory within the court hierarchy. The following set of labels indicates a transition from one state to another: \texttt{Charges Accepted}, \texttt{Charges Refused}, \texttt{Pled Guilty}, \texttt{Pled Not Guilty}, \texttt{Sentence with Incarceration Imposed}, \texttt{Sentence Without Incarceration Imposed}, \texttt{Found Not Guilty}.

However, several methodological challenges persist at this stage. A primary concern, even following successful label extraction, pertains to the significant degree of noise and procedural irregularities encountered throughout a case's lifespan. Furthermore, we must account for entries that, despite containing substantive data captured by our extraction algorithm, occur in parallel to the formal procedure. This phenomenon stems from the decentralized nature of the court system. For instance, a case may be actively prosecuted within a criminal district court (\emph{CDC}), yet continue to generate concurrent entries at the magistrate level (\emph{MAG}), complicating the chronological mapping of the judicial flow.

Another significant challenge involves procedural inconsistencies, specifically the omission or lack of formal documentation for certain judicial steps. For instance, a case may transition abruptly from \emph{First appearance} to the \emph{Trial process} via a \texttt{Pled Not Guilty} label, without any preceding record indicating that \texttt{Charges Accepted} has occurred. Similarly, a proceeding text might imply that a trial has reached a verdict, yet lacks a subsequent entry confirming the sentencing and its specific classification.

Furthermore, it is not uncommon for multiple legal states to be bypassed within a single proceeding. Complexities also arise with labels such as \texttt{Pled Guilty}, which may manifest at various stages of the litigation, necessitating the observation of a nuanced chain of events rather than a fixed chronological point. Finally, cases involving multiple charges present a unique difficulty; while a specific charge might be refused, others are accepted and pursued. This results in datasets containing proceedings that reference both labels, either simultaneously or in serial, thereby complicating the linear modeling of the case flow.

To monitor the progression of case states, we employ the graph structure in \figref{flowchart}. Each state is assigned a comprehensive list of potential labels representing all outgoing edges on a path toward \emph{Case Closed}. For instance, at the \emph{Pre-arraignment} stage, the permissible labels include \texttt{Pled Not Guilty}, \texttt{Pled Guilty}, \texttt{Sentence with Incarceration Imposed}, \texttt{Sentence Without Incarceration Imposed}, and \texttt{Found Not Guilty}. Conversely, at the \emph{Trail process} stage, the subset is restricted to the latter three labels.

The implementation of our tracking system follows a specific chronological logic: for every case, proceedings are ordered by date; in instances where two proceedings occur on the same day, the original sequence is maintained. Commencing at the top of this ordered list, we identify the initial state associated with the first entry (\emph{First appearance}). At each subsequent state, the system parses forward to identify a label corresponding to the list of possible outcomes. Once a label is identified, the case's position advances within the graph; this iterative process continues until the case reaches \emph{case closed} or the sequence of proceedings is exhausted.

One complication arises since, as previously noted, a case may be refused in the criminal district court (\emph{CDC}) at an advanced stage, even after initial acceptance at the magistrate level. Additionally, there is the possibility that specific charges are refused while others remain active. Consequently, care is required to prevent the premature termination of a case or, conversely, leaving a case designated as ``open'' when it has reached a functional conclusion. To mitigate these risks, we only interpret \texttt{Charges Refused} as a definitive signal for case closure if it is not followed by any subsequent proceedings containing valid edge labels.

By employing this methodology, we generate a comprehensive timeline for each case, detailing the specific sequences of states traversed throughout the judicial process. This structured data subsequently serves as the foundation for analytical plots designed to visualize the duration of each distinct state and their respective final outcomes.

\subsection{Detecting Pre-trial Detainment}
\label{sec:pretrial}
Pre-trial detainment (whether a defendant remains incarcerated between arrest
and case resolution) is a central transparency question for New Orleans
stakeholders \cite{armstrong2022unconvicted,armstrong2022louisiana}, yet it is not recorded as a structured field in Docket Master.
Clerks instead leave sporadic linguistic cues across chronological minutes:
references to custody, jail lists, transport failures, bond posting, and
release on recognizance.
Our goal is therefore to reconstruct dated custody intervals from these cues
and to mark which intervals fall before adjudication.

\paragraph{Custody-relevant labels}
Several labels in $\mathcal{L}$ supply the raw signals for this task.
\texttt{Defendant in Custody and Not Present in Court} captures entries in
which the defendant is reported as incarcerated and was not produced for the
listed hearing. Because clerks typically name specific institutions rather than using generic custody language, the corresponding extractor embeds lists of Louisiana
correctional facilities and parishes into its compiled patterns, and requires
a concurrent non-appearance match (e.g., not brought, not transported, or
absent) to reduce false positives from administrative wording alone.
\texttt{Released on Own Recognizance} identifies affirmative release language
while applying exclusion rules for denials and counterfactual phrasing
(e.g., not eligible, rescinded).
\texttt{Bond Set} and \texttt{Bond Not Set} record bond disposition, and the
sentencing extractors distinguish
\texttt{Sentence with Incarceration Imposed} from
\texttt{Sentence Without Incarceration Imposed}, which later helps separate
post-adjudication jail time from pre-adjudication detention.

\paragraph{Classifying each proceeding}
Beyond the structured labels, we classify each proceeding's free text into one
of three custody statuses: \emph{in custody}, \emph{released}, or
\emph{unknown}.
In-custody phrases include mentions of being in custody or federal custody,
placement on a jail list, transport to court, arrest language, and related
administrative markers.
Release phrases include posted bond or bail, surety, house arrest, release
issued, and non-appearance signals that typically accompany a defendant who
is not currently held (e.g., capias issued, failed to appear).
If a proceeding already carries the \texttt{Released on Own Recognizance}
label, we force its status to \emph{released}, overriding ambiguous text.

\paragraph{Building custody intervals}
For every case, we order proceedings by date while preserving the original order
when two entries share a day and walk this sequence chronologically, as in our case-state tracker.
When we encounter an \emph{in custody} entry, we open a custody interval and
record that entry's date as the start.
The next \emph{released} entry for the same case closes the interval; its date
becomes the end, and the duration is the difference between the two dates.
If a new case begins while a interval is still open, we close the open interval at
the last date of the preceding case rather than allowing custody intervals to
cross case boundaries.
Proceedings whose status is unknown leave the current interval unchanged.
The result is a sparse set of custody intervals per magistrate number, each
with an explicit start date, end date, and length in days.

\paragraph{Marking intervals as pre-trial}
Not every custody interval is pre-trial: defendants may return to custody
after a plea, verdict, or sentence.
To isolate pre-adjudication detention, we first locate each case's earliest
\emph{resolution date}, the first proceeding labeled
\texttt{Charges Refused}, \texttt{Guilty Plea Entered},
\texttt{Found Guilty}, or \texttt{Found Not Guilty}.
A custody interval is flagged as pre-trial detainment when this resolution date
falls within the interval $[\textit{Start Custody},\textit{End Custody}]$
(or when the end date is missing but the start precedes resolution).
Cases with at least one such interval receive a case-level pre-trial detainment
indicator.
For visualization and case-detail views, we additionally stamp the opening
proceeding of each pre-trial interval with \texttt{Custody Starts} and the
closing proceeding with \texttt{Custody Ends}.

This procedure yields dated start and end times without parsing clock times
from free text: timing is inherited from the structured proceeding dates
already present in the docket.
Combined with the label library in Section~\ref{sec:label}, it converts noisy
clerk language into custody intervals that the dashboard can filter,
aggregate, and display alongside bond amounts, demographics, and case-flow
outcomes.

\subsection{Finding Similar Cases}
\label{sec:similar}

%  using the Case Details view often ask not only what occurred in a single docket, 
Stakeholders often want to know not only what occurred in a single case, but also which completed cases are comparable in terms of charges, defendant demographics, bond amounts, magistrate sections, and clerk language. Because Docket Master does not expose such nearest neighbors, we embed each case as a single document and retrieve similar cases in that embedding space, enabling the dashboard to list comparable records alongside a selected magistrate number.

\paragraph{Case-as-document representation}
Each magistrate number is treated as a single document by concatenating, in chronological order, tokenized representations of every proceeding belonging to that case. These tokens integrate structured fields including proceeding dates, original and total bond amounts, race, gender, age, magistrate sections, charge counts, and individual statutes, with free text from the clerk's minute notes. Record identifiers, label strings, and other bookkeeping attributes are omitted so that similarity is driven by substantive case characteristics rather than internal keys. Charges formatted as pipe-separated lists are split into individual statute tokens, while numerical race and gender codes are converted into concise lexical labels (e.g., Black, female). Additionally, minute text is lowercased and reduced to alphanumeric tokens. Every token is prefixed with its field name (e.g., \texttt{charge=}, \texttt{race=}, \texttt{text=}), allowing the model to distinguish bond amounts from ages, and statutory charge strings from general words in clerk notes.

\paragraph{Embedding}
We train a distributed-memory Doc2Vec \cite{LeMikolov2014} model on this corpus, where each case receives a 100-dimensional vector updated over 20 epochs to predict document tokens from local context alongside the case's own representation. Because documents vary considerably in length, paragraph embeddings offer a better fit than fixed-schema feature vectors; although longer cases contribute more tokens, all cases are projected into the same unified comparison space. The resulting vectors are stored for reuse and recomputed only when the underlying case tables are updated.\footnote{Note that this embedding is a first pass and does not consider potential bias or weighting for legally relevant concepts; we are exploring alternative ranking methodologies for future work.} 

\begin{figure}[t]
    \centering
    \includegraphics[width=\linewidth]{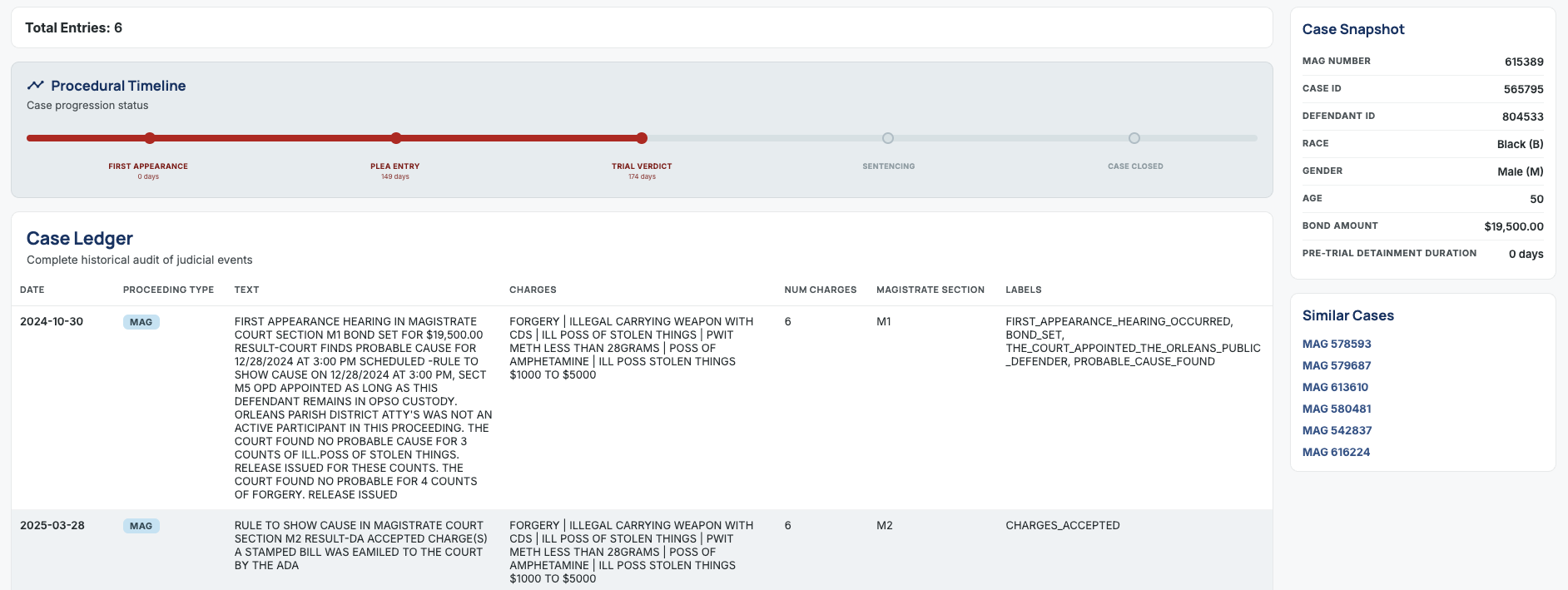}
    \caption{Case details page for an open case, containing case timeline, key information, similar cases and proceedings.}
    \label{fig:casedetails}
\end{figure}

\paragraph{Retrieval}
Given a query magistrate number, we score all other cases using the cosine distance between embedding vectors ($1 - \text{cosine similarity}$), where $0$ indicates identical orientation, and higher values signify greater dissimilarity. Cases are ranked from nearest to farthest, allowing the dashboard to display either the complete ranking or the top nearest neighbors. This neighbor list populates the similar-cases panel within the Case Details view (\figref{casedetails}, offering users a concrete set of comparators when evaluating delays, bond amounts, or outcomes for a specific docket.

\subsection{Visualizations and Data Presentation}
\label{sec:vis}
\begin{figure}[ht!]
    \centering
    \includegraphics[width=\linewidth]{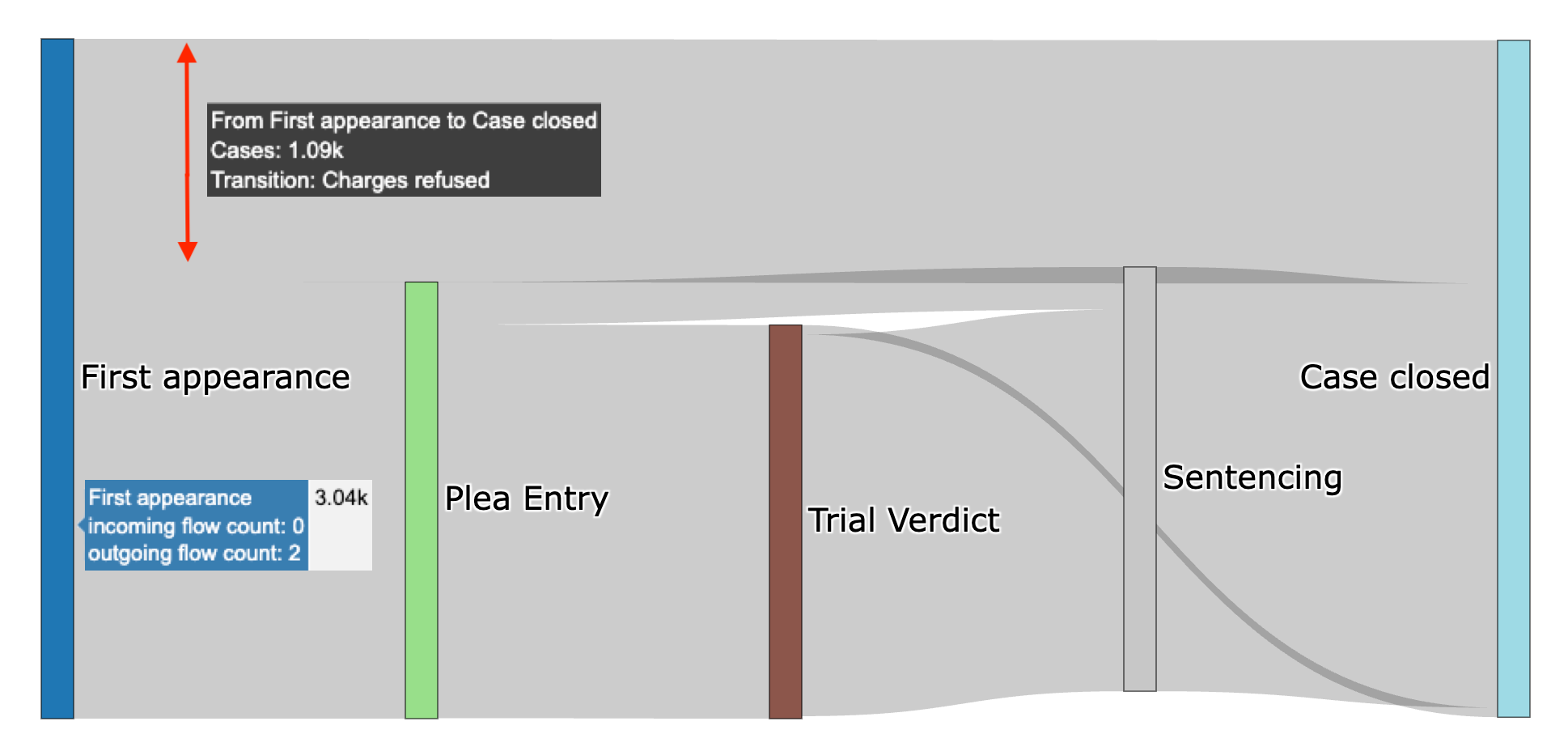}
    \caption{Case flow diagram showing charges dropped for 35.85\% of the cases corresponding to a male defendant in 2024 and 2025. The popup boxes appear upon hovering over the flow edges providing more detail.
    }
    \label{fig:men-chargedrop}
\end{figure}

\begin{figure}[ht!]
    \centering
    \includegraphics[width=\linewidth]{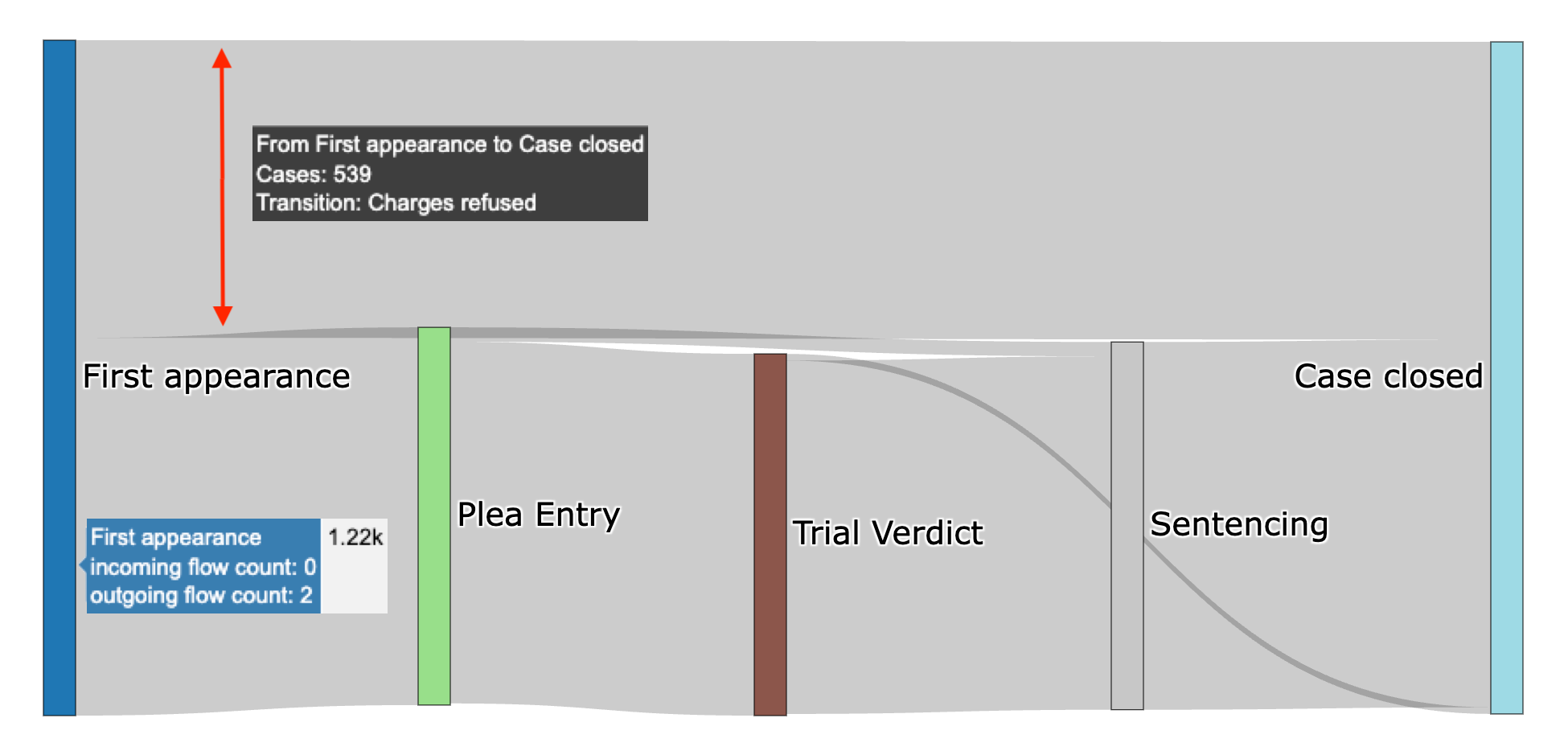}
    \caption{Case flow diagram showing charges dropped for 44.07\% of the cases corresponding to a female defendant in 2024 and 2025.The popup boxes appear upon hovering over the flow edges providing more detail.}
    \label{fig:women-chargedrop}
\end{figure}

Our visualizations are designed to meet the needs expressed both in a sequence of community meetings held as part of the design and development of the system, as well as drawing on other approaches in the literature \cite{mcc2024nola,TurnerJohnson2006,CCJ2024WomensJustice,QuattroneCenter2017,stacked2023,vera2017nola}. Our community partners include law and criminology faculty, researchers, criminal justice and crime survivor advocates, legal professionals, investigative journalists, and members of non-governmental organizations (e.g., \url{https://www.courtwatchnola.org/}). These plots specifically highlight key factors commonly studied by experts in the field. Through this process, we have developed six different visualizations, each designed to address a specific need.
\begin{itemize}[nosep]
    \item \textbf{What happened in court}: scatter plot of all proceedings by calendar date vs. days since first entry, useful for identifying case delays over time. (e.g., \figref{whathappened} and \figref{m3}) %\todo{update each figure with this title or point this title to each figure?}
    \item \textbf{Case flow}: a Sankey diagram showing transitions between states; thickness reflects case volume; useful for identifying aggregate trends in case flows. (e.g., \figref{men-chargedrop} and \figref{women-chargedrop})
    \item \textbf{Progression by label}: horizontal timelines of labeled segments per case; useful for comparing case event sequences. (e.g., \figref{custody})
    \item \textbf{Progression by state}: horizontal timelines of state per case; useful for comparing state sequences. (e.g., \figref{black} and \figref{white})
    \item \textbf{Bond distribution}: histogram, (e.g., \figref{whitebond} and \figref{blackbond})
    \item \textbf{Case Ledger (Top-5 and Bottom-5 cases)}: a table listing the top or bottom 5 cases by duration, first appearance date, pre-trial detainment duration, and bond amount.
    \item \textbf{Case Details}: Containing a case timeline diagram, critical information about the case, proceedings and a list of similar cases. (e.g., \figref{casedetails})
\end{itemize}

To get visualization tailored to a specific question, we offer a variety of data filters, including date range, demographics, magistrate section, charges, and case outcomes. In the following, we provide some examples of these visuals and the studies that helped shape our visualizations.

%Our visualizations are designed to meet the community's needs, drawing on previous research and articles in criminal justice. These plots specifically highlight key factors commonly studied by experts in the field. In the following, we provide several examples of these studies and how they helped shape our visualizations.

\subsubsection{Unusual Delays}
\begin{figure}[t]
    \centering
    \includegraphics[width=\linewidth]{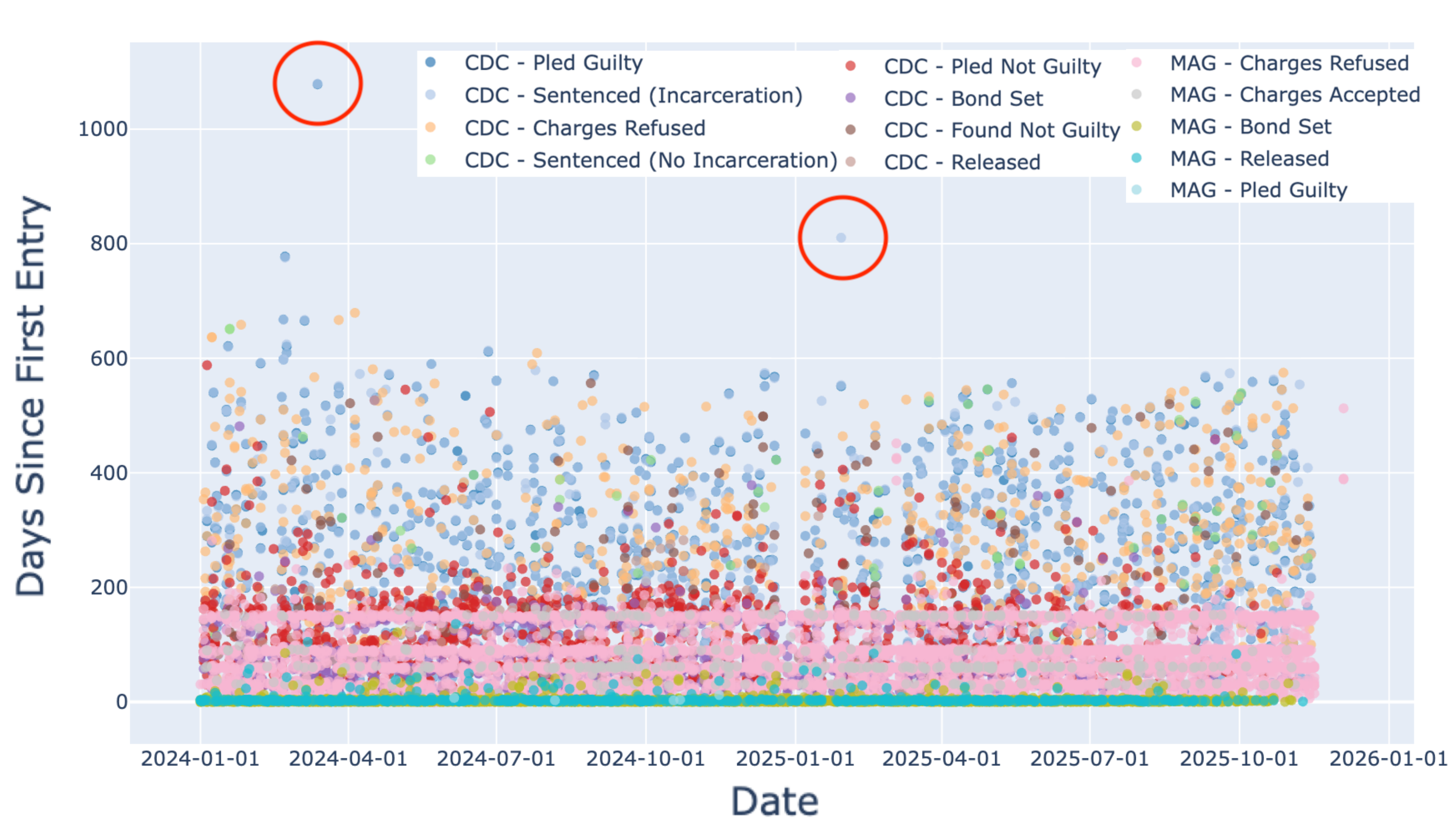}
    \caption{All proceedings in all sections in 2024 and 2025. Each point shows a proceeding and each color specifies a corresponding label which can be selected or deselected by the user.}
    \label{fig:whathappened}
\end{figure}

\begin{figure}[t]
    \centering
    \includegraphics[width=\linewidth]{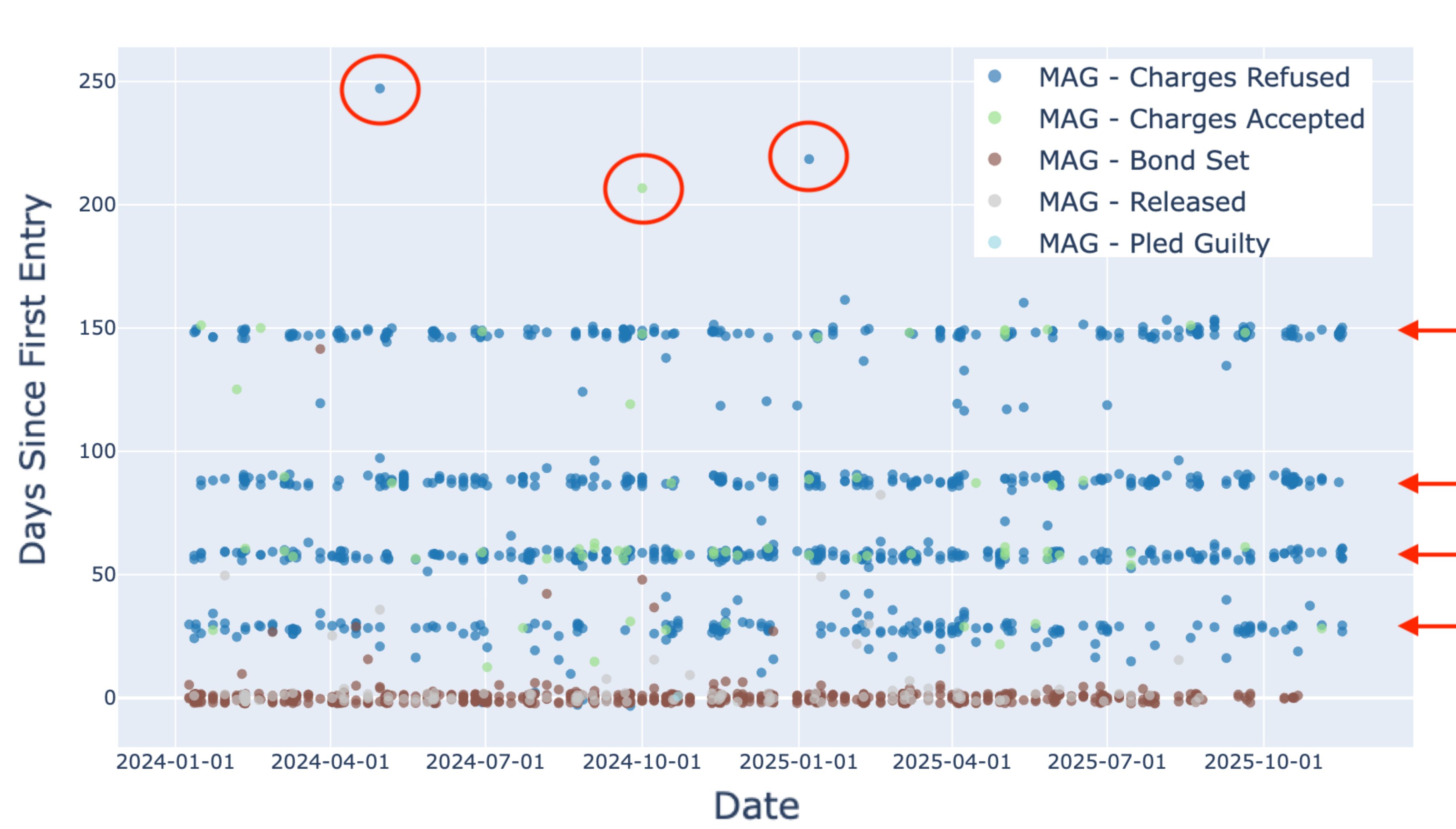}
    \caption{All proceedings in one section of Magistrate court in 2024 and 2025. Each point shows a proceeding and each color specifies a corresponding label which can be selected or deselected by the user.}
    \label{fig:m3}
\end{figure}

% making this less judge adversarial -awc
Timeline disparities remain a pervasive issue in the judicial system, with significant variations in case management for reasons including case complexity, evidence collection, and caseloads of court sections~\cite{mcc2024nola}.
% Prior work has found that case resolution varies significantly by court section, creating a "geographic lottery" of justice
 To surface this variation and facilitate assessment, we mapped labeled proceedings onto a coordinate grid defined by calendar date and elapsed duration since \emph{First appearance}. This visualization can be refined by applying a section filter to isolate trends by court section. As shown in \figref{whathappened}, some cases received updates after $600$ days (red circles). Additionally, \figref{m3} shows updates in one section of Magistrate court, indicating decisions are made close to legal deadlines (red arrows). Article 701 of the Louisiana Code of Criminal Procedure mandates strict timelines (varying by charge type) for bringing a detained person to trial or formally charging them.
% Disparities in timelines are one of the most common problems in the judicial system, often causing confusion and frustration. For instance, there are wide disparities in how different judges handle cases. An article~\cite{mcc2024nola} shows that while some sections rushed to reduce their median case resolution time to $195$ days, others allowed cases to languish for over a year, creating a "geographic lottery" of justice within the same building. To highlight this variation and facilitate a more granular assessment of each section's performance, we mapped all labeled entries onto a coordinate grid defined by the calendar date of the entry and the elapsed duration in days since the \emph{First appearance}. This visualization can be further refined by applying a specific section filter to isolate departmental trends. As shown in \figref{whathappened}, certain cases continued to receive updates even after a period of $600$ days, although such occurrences have become significantly less frequent throughout 2025. Additionally, \figref{m3} shows the pattern of updates in the magistrate court section 3, indicating that decisions are often made very close to the deadlines.


\subsubsection{Gender and Conviction Rates}\label{sec:gender}
Existing research indicates that women are significantly less likely to proceed to trial \cite{Starr2012} and exhibit a $20\%$ higher probability of having all charges dismissed \cite{TurnerJohnson2006,CCJ2024WomensJustice}. These disparities are reflected in our case flow diagrams examining $3,038$ unique male and $1,223$ unique female cases. For the male cohort, $35.85\%$ of cases resulted in dropped charges, while $64.15\%$ were formally accepted. Of continued proceedings, $90.25\%$ involved a plea of not guilty or transition to trial, yielding a $97.15\%$ conviction rate compared to a $2.85\%$ acquittal rate. In contrast, female cases revealed $44.07\%$ resulted in dropped charges, with $55.93\%$ accepted for prosecution. For proceeded cases, $95\%$ entered a plea of not guilty or moved to trial, concluding with a $97.71\%$ guilty verdict and a $2.29\%$ not-guilty verdict. Comparative trajectories are illustrated in \figref{men-chargedrop} and \figref{women-chargedrop}.

% Existing research indicates that women are significantly less likely to proceed to trial \cite{Starr2012} and exhibit a $20\%$ higher probability of having all charges dismissed \cite{TurnerJohnson2006,CCJ2024WomensJustice}. These disparities are also reflected in our case flow diagrams, which examined $3038$ unique male cases and $1223$ unique female cases. Regarding the male cohort, $35.85\%$ of cases resulted in charges being dropped, while $64.15\%$ were formally accepted. Of the proceedings that continued, $90.25\%$ involved either a plea of not guilty or a transition to trial, ultimately yielding a $97.15\%$ conviction rate compared to a $2.85\%$ acquittal rate.

% In contrast, the analysis of female cases revealed that $44.07\%$ resulted in dropped charges, with $55.93\%$ being accepted for prosecution. For the subset of cases that proceeded, $95\%$ entered a plea of not guilty or moved to trial; these cases concluded with a $97.71\%$ guilty verdict and a $2.29\%$ not-guilty verdict. These comparative trajectories are further illustrated in \figref{men-chargedrop} and \figref{women-chargedrop}.
\subsubsection{Race and Longer Procedures}\label{sec:race}
\begin{figure}[t]
    \centering
    \includegraphics[width=\linewidth]{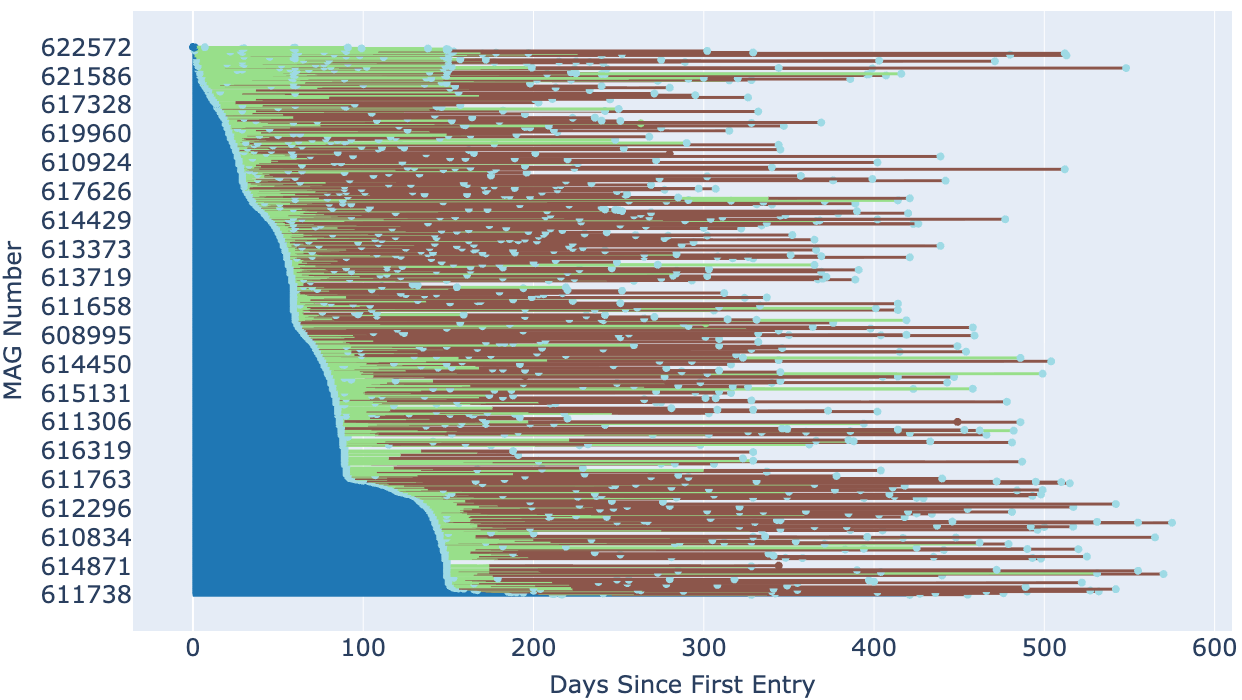}
    \caption{Case progression of cases where the defendant(s) is black. Colors show the same states as Figure \ref{fig:women-chargedrop}.}
    \label{fig:black}
\end{figure}

\begin{figure}[t]
    \centering
    \includegraphics[width=\linewidth]{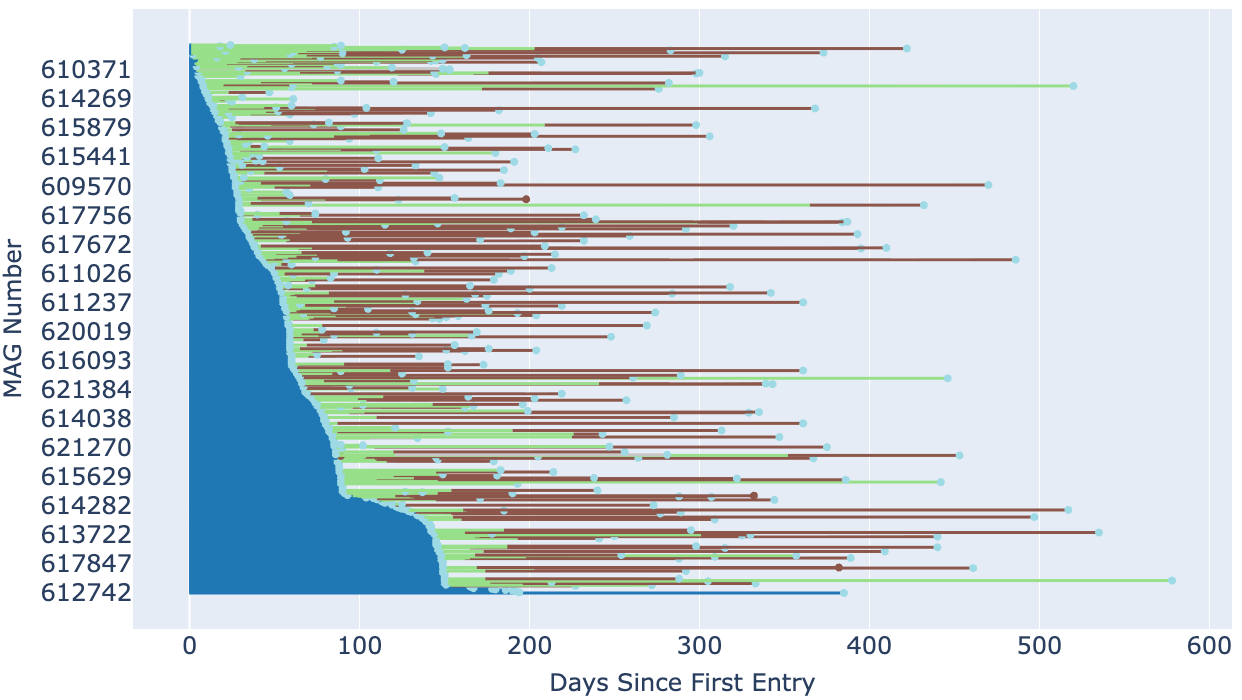}
    \caption{Case progression of cases where the defendant(s) is white. Colors show the same states as Figure \ref{fig:women-chargedrop}.}
    \label{fig:white}
\end{figure}

As anticipated, case progression varies significantly by race~\cite{QuattroneCenter2017}. A comparative analysis of \figref{black} and \figref{white} reveals that Black defendants experience protracted durations before case updates, remaining in a pending status within each judicial state for longer periods. Specifically, there is a higher density of cases reaching the $500$--$600$ day range. Furthermore, cases involving Black defendants exhibit approximately $16.1\%$ longer durations than white defendants, with the average duration extending from $77.5$ days to $90$ days. However, observed trends in \secref{gender} and \secref{race} may be a function of the underlying nature of charges and specific types of offenses attributed to each demographic group. These confounding variables require deeper analysis to isolate the exact causes of procedural gaps; our dashboard provides the necessary tools for such an investigation.

% As anticipated, the defendant's race significantly influences case progression \cite{QuattroneCenter2017}. A comparative analysis of \figref{black} and \figref{white} reveals that Black defendants experience protracted durations before receiving case updates, typically remaining in a pending status within each judicial state for longer periods. Specifically, there is a higher density of cases reaching the $500$--$600$ day range for this demographic. Furthermore, cases involving Black defendants exhibit an approximate $16.1\%$ temporal delay relative to those of white defendants, with the average duration extending from $77.5$ days to $90$ days. 

% However, we must acknowledge that the observed trends discussed in \secref{gender} \secref{race} may be a function of the underlying nature of the charges and the specific types of offenses attributed to each demographic group. These confounding variables require deeper analysis to isolate the exact causes of procedural gaps; our dashboard provides the necessary tools for this type of investigation.

\subsubsection{Criminal Charge Stacking} 

\begin{figure}[t]
    \centering
    \includegraphics[width=\linewidth]{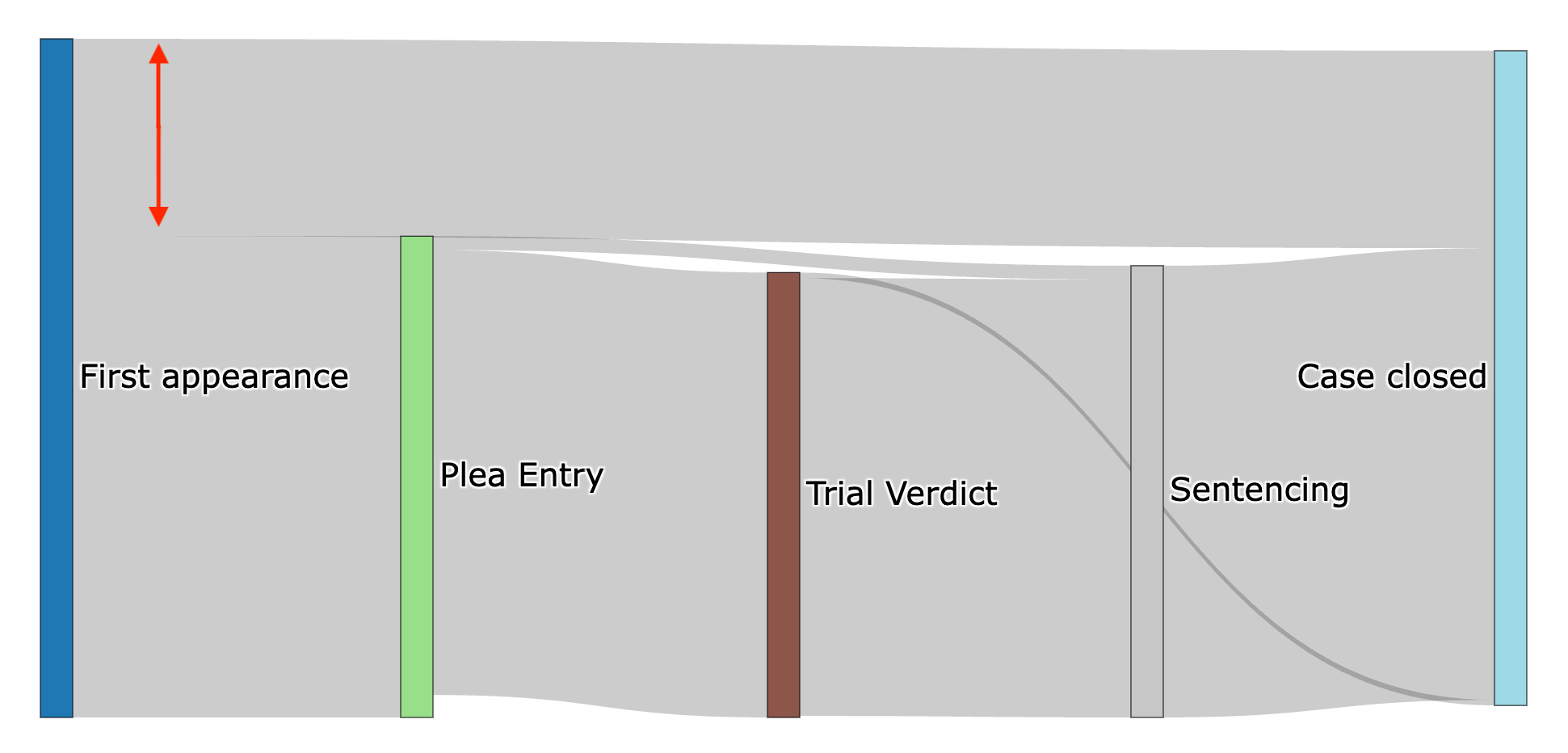}
    \caption{A case flow plot showing cases with one charge.}
    \label{fig:one-charge}
\end{figure}

\begin{figure}[t]
    \centering
    \includegraphics[width=\linewidth]{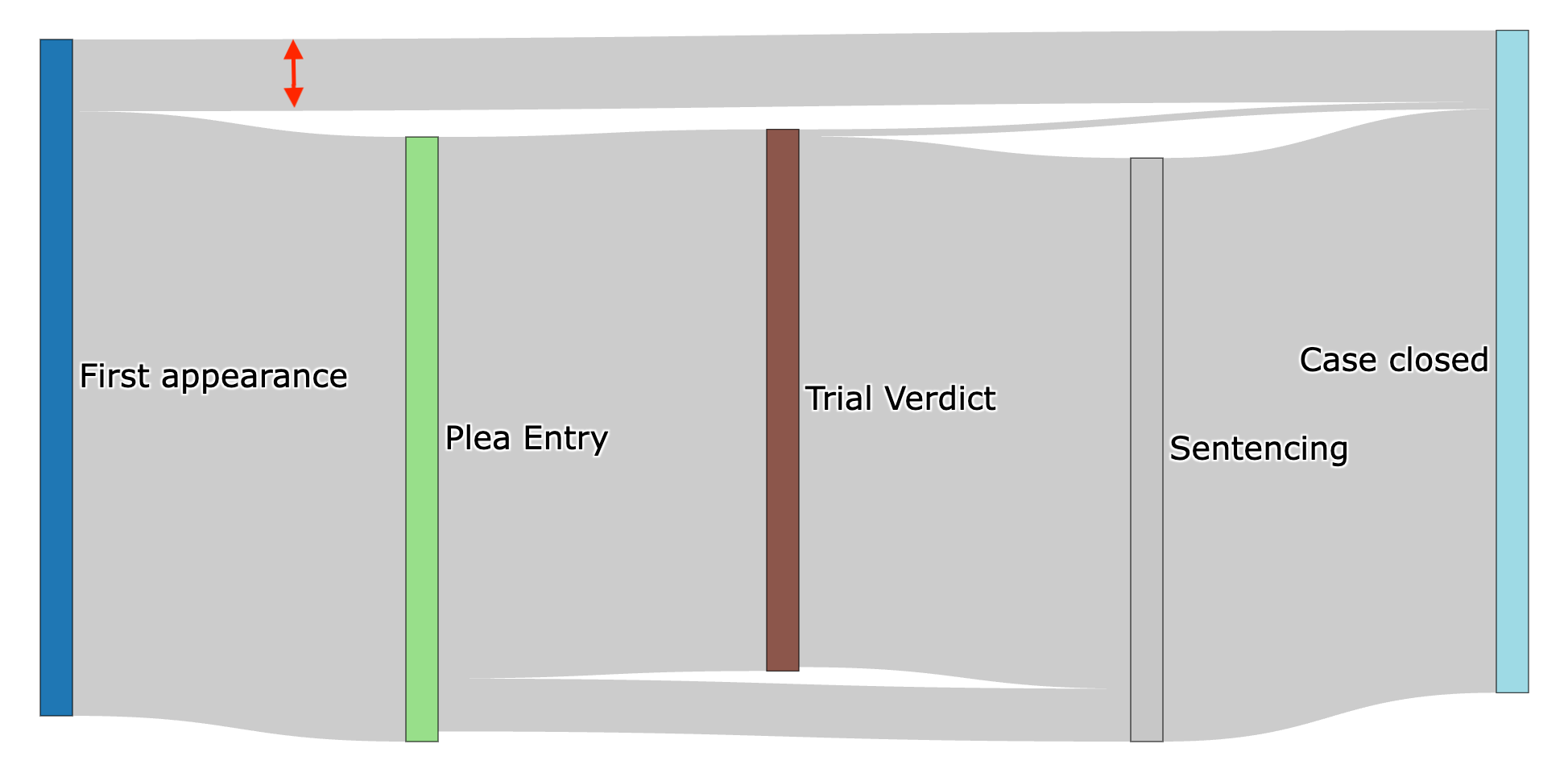}
    \caption{A case flow plot showing cases with two or more charges.}
    \label{fig:two-charge}
\end{figure}

Our empirical case flow visualizations validate existing literature on judicial charging decisions (\figref{one-charge} and \figref{two-charge}). Charge stacking techniques, such as filing multiple and overlapping charges to maximize potential sentencing and pressure defendants \cite{dholakia2024criminal,crespo2018hidden}, have been criticized for forcing individuals into plea bargains or undermining success at trial \cite{neily2019prisons}. Federal court mostly charges defendants with a single offense \cite{stacked2023}; in contrast, state justice systems levy multiple charges far more frequently.

Our data shows an increased number of charges correlates directly with more adverse outcomes for defendants. 
Looking at primary final dispositions across $8,923$ single-charge cases (\figref{one-charge}), $27.1\%$ were dropped, $60.0\%$ resulted in a guilty verdict, and $0.7\%$ ended in acquittal. Along the procedural timeline, $66.0\%$ of all cases involved a plea entry and $61.0\%$ proceeded to trial. Among $5,737$ dual-charge cases (\figref{two-charge}, primary dispositions shifted significantly: dismissals fell to $9.0\%$, while convictions and acquittals accounted for $72.1\%$ and $0.9\%$ of total cases, respectively. Procedurally, $82.0\%$ involved pleas and $73.5\%$ went to trial.

\subsubsection{Bond Discrepancies}

%There is a significant body of reporting and academic research regarding bond discrepancies and the opacity of the bail system, with New Orleans often serving as a primary case study due to its unique funding structure, where the lack of transparency has historically been tied to a direct conflict of interest. Until relatively recently, a portion of every bond fee paid in Orleans Parish went directly into a Judicial Expense Fund used by judges to pay for staff, travel, and office supplies~\cite{vera2017nola}.

\begin{figure}[t]
    \centering
    \includegraphics[width=\linewidth]{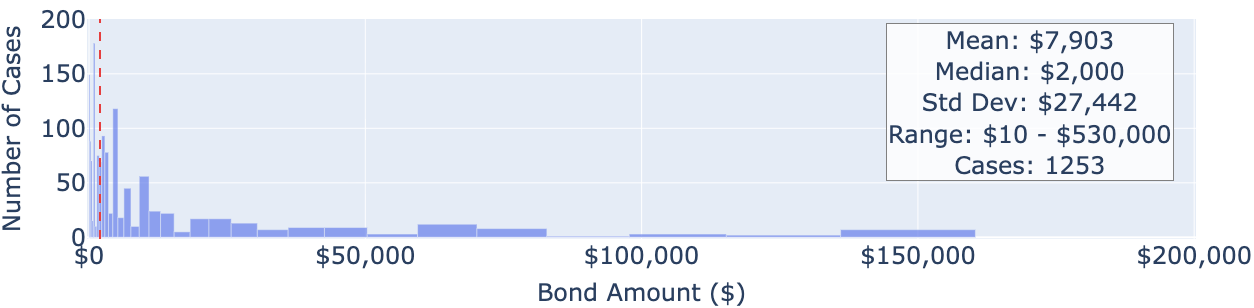}
    \caption{Bond distribution for battery charges by White defendants.}
    \label{fig:whitebond}
\end{figure}

\begin{figure}[t]
    \centering
    \includegraphics[width=\linewidth]{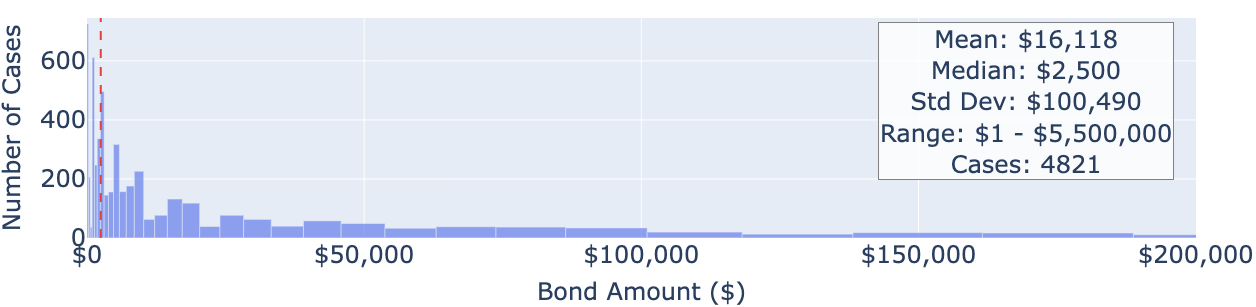}
    \caption{Bond distribution for battery charges by Black defendants.}
    \label{fig:blackbond}
\end{figure}

There is a significant body of reporting and academic research regarding bond discrepancies and the opacity of the bail system, with New Orleans often serving as a primary case study due to its unique historical funding structure.
%, where the lack of transparency has historically been tied to a direct conflict of interest. 
Until relatively recently, a portion of every bond fee paid in Orleans Parish went directly into a Judicial Expense Fund used by judges to pay for staff, travel, and office supplies~\cite{vera2017nola}. Our dashboard offers statistics on bond amounts based on our filters. For example, \figref{whitebond} and \figref{blackbond} show the bond amounts for battery charges based on race, indicating that while the typical bond amounts are similar for both groups, black defendants are subject to a wider range of exceptionally high outlier bonds that significantly inflate their overall average. Again, additional case context (charges, history, etc.) is required to interpret these findings.


\subsubsection{Pre-trial Detainment Duration}
\begin{figure}[t]
    \centering
    \includegraphics[width=\linewidth]{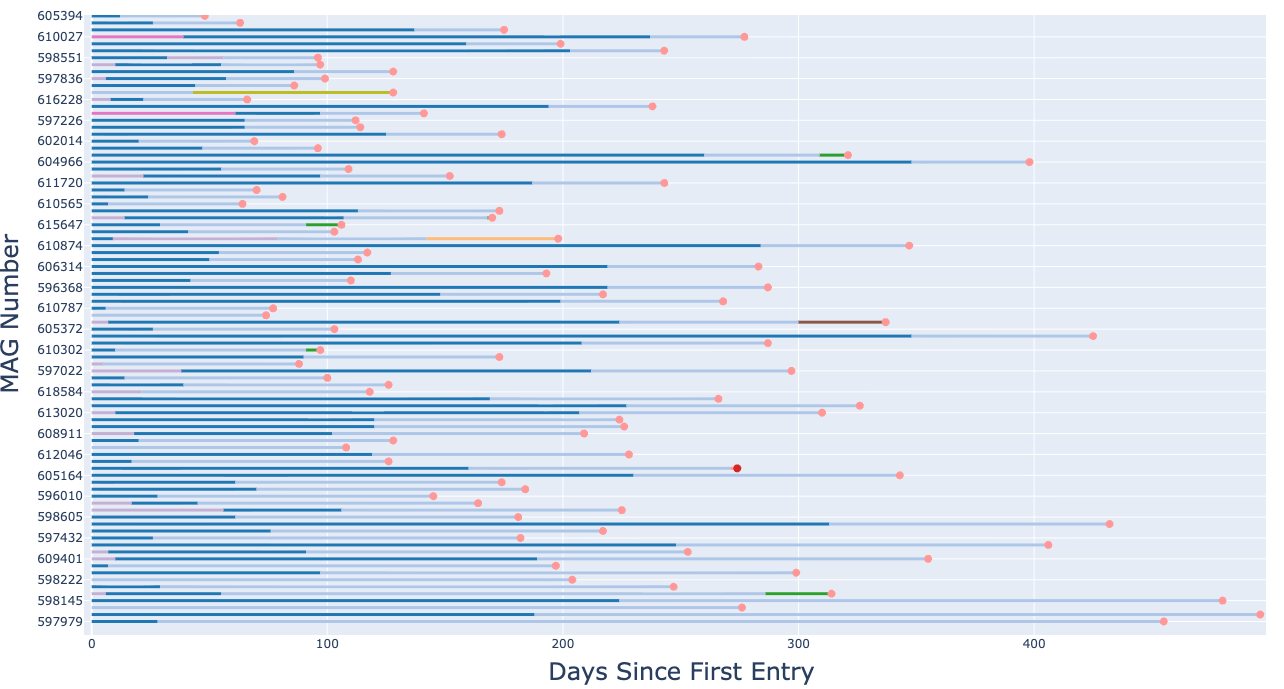}
    \caption{Case progression plot for cases with longest pre-trial detainment between 2022 and 2026. Light blue shows custody length in days. Orange dots show end of custody.}
    \label{fig:custody}
\end{figure}

Pre-trial detainment serves as a deterrent to defendant flight risk and to protect future harm to victims, but can also be a result of high bond requirements. This can also serve as a leverage point to encourage plea deals and avoid lengthy trials~\cite{armstrong2022louisiana,lattimore2021reimagining}. Supporting this, our visualizations demonstrate that $77.4\%$ of detained defendants plead guilty from 2022 to 2026, including cases where individuals remained in custody for up to $400$ days (\figref{custody}) whereas only $56.8\%$ of cases end with a guilty plea where the defendant was not detained. Moreover, our analysis identifies possible Article 701 violations which set the strict time limits for when the state must file formal charges, hold a preliminary examination, and commence trial after a defendant is arrested ($30$ days for misdemeanors in custody, $60$ days for felonies). Finally, bond amounts for pre-trial detainees show an unusual pattern; rather than tapering off smoothly at higher values, the distribution features several distinct peaks specifically at high amounts.

\section{Formative Study}
%\aron{talk about the previous (informal) community workshops when we asked the partners what they wanted}
As we showed in \secref{vis}, our system can help surface existing claims found across law and sociology literature, investigative journalism, and governmental reports. This illustrates the tool's capability to assist researchers and authors in monitoring and discovering anomalies.
%To validate our visualization designs, we first demonstrated in \secref{vis} that the system accurately surfaces existing claims found across law and sociology literature, investigative journalism, and governmental reports. By establishing this alignment with recognized documentary evidence, we illustrate the tool's capability to assist researchers and authors in discovering novel trends and systemic anomalies.
To complement this, we conducted a small formative study in collaboration with local stakeholders and our community partners.
%This cohort included law and criminology faculty, researchers, criminal justice and crime survivor advocates, legal professionals, investigative journalists, and partner members of non-governmental organizations (e.g., \url{https://www.courtwatchnola.org/}). \aron{move this to \secref{vis}}
Participants were initially presented with a system demonstration featuring representative queries. Following this, they were provided independent access to the dashboard and subsequently submitted feedback via a structured survey. While preliminary, participants generally reported positive experiences regarding the system's usability and utility. Specifically, participants commended the intuitive design of the case progression plot, while others noted that the granular presentation of case details clarified broader legal trajectories and critical judicial events. One suggestion noted that while the level of detail the dashboard provides is useful, simpler visualizations may be more appropriate for broader audiences.

\section{Conclusion and Future Work}
We present a data management and visualization system designed in collaboration with community partners to enhance transparency and public oversight within the criminal court system. By providing specialized analytical tools to researchers, policymakers, legal practitioners, and crime survivors, the platform facilitates the tracking and discovery of systemic trends, anomalies, and policy-driven impacts. The system effectively highlights established phenomena within case flows, timelines, and legal outcomes while simultaneously offering the exploratory capabilities necessary to identify understudied patterns. 
%The design and implementation of these tools were informed by a formative study conducted in collaboration with interdisciplinary partners across criminal law, sociology, and survivor advocacy.

While the current iteration of the dashboard offers significant utility, several avenues for future development have been identified and suggested by our partners. Future versions will integrate other sources such as police reports and incorporate more case specific information such as evidence and witnesses. %Future versions will integrate the identification and visualization of pre-trial detention periods, enabling an empirical analysis of how jail time correlates with case trajectories and final outcomes.
Furthermore, we aim to implement a more robust empirical similarity measure derived from our findings, to group analogous cases and streamline the detection of broader judicial trends. %clustering feature using an 
Ultimately, we anticipate that this dashboard will serve as a catalyst for increasing public trust and institutional accountability by ensuring that judicial decision-making processes remain accessible and transparent.

%% if specified like this the section will be omitted in review mode

\section*{Acknowledgment}
We thank our local and community partners, Court Watch NOLA, Jesse Manley, Andrea Armstrong, Darrin Browder, Andrea Boyles, and Julie Ford, for the fruitful talks and feedback.
E. Hosseini Sereshgi was supported by the Murphy Institute;
	N. Mattei was supported in part by NSF Awards IIS-RI-2339880, CNS-SCC-2427237, and IIS-HCC-2619509; A. Culotta was supported in part by NSF Awards IIS-HCC-2619509, DEB-2550002, CNS-2515250, IIS-HCC-2333537, CNS-2427237, the Louisiana Board of Regents Endowed Chairs for Eminent Scholars program, and the Greenberg Family Professorship in Social Entrepreneurship. The research was additionally supported by the Harold L. and Heather E. Jurist Center of Excellence for Artificial Intelligence at Tulane University and the Tulane University Center of Excellence for Community-Engaged Artificial Intelligence (CEAI).

\bibliographystyle{abbrv-doi-hyperref}

\bibliography{references}

%\appendix % You can use the `hideappendix` class option to skip everything after \appendix
%\crefalias{section}{appendix} % this is to make sure that cleverref switches to referring to Appx. X from here on

%\section{About Appendices}
%Refer to \cref{sec:appendices_inst} for instructions regarding appendices.

\end{document}